\documentclass[aps,pre,showpacs,twocolumn,superscriptaddress]{revtex4}
\usepackage{graphicx,color,epsfig}
\providecommand{\e}[1]{\ensuremath{\times 10^{#1}}}

\begin{document}


\title{Consensus and ordering in language dynamics}

\author{Xavier Castell\'o}
\affiliation{IFISC, Institut de F\'isica Interdisciplin\`aria i Sistemes
  Complexos (CSIC-UIB), Campus Universitat Illes Balears E07122 Palma
  de Mallorca (Spain)}
\author{Andrea Baronchelli}
\affiliation{Departament de F{\'i}sica i Enginyeria
  Nuclear, Universitat Polit{\`e}cnica de Catalunya, \\ Campus Nord B4,
  08034 Barcelona (Spain)}
\author{Vittorio Loreto}
\affiliation{Dipartimento di Fisica, ``Sapienza''
  Universit\`a di Roma, P.le A. Moro 2, 00185 Roma (Italy)}
\affiliation{Fondazione ISI, Viale S. Severo 65, 10133 Torino (Italy)}

\date{\today}

\begin{abstract}
We consider two social consensus models, the AB-model and
  the Naming Game restricted to two conventions, which describe a
  population of interacting agents that can be in either of two
  equivalent states (A or B) or in a third mixed (AB) state. Proposed
  in the context of language competition and emergence, the AB state
  was associated with bilingualism and synonymy respectively. We show
  that the two models are equivalent in the mean field approximation,
  though the differences at the microscopic level have non-trivial
  consequences. To point them out, we investigate an extension of
  these dynamics in which confidence/trust is considered, focusing on
  the case of an underlying fully connected graph, and we show that
  the consensus-polarization phase transition taking place in the
  Naming Game is not observed in the AB model. We then consider the
  interface motion in regular lattices. Qualitatively, both models
  show the same behavior: a diffusive interface motion in a
  one-dimensional lattice, and a curvature driven dynamics with
  diffusing stripe-like metastable states in a two-dimensional
  one. However, in comparison to the Naming Game, the AB-model
  dynamics is shown to slow down the diffusion of such configurations.
\end{abstract}


\maketitle

\section{Introduction}
\label{intro}

The formalism, ideas and tools from statistical physics and complex
systems have successfully been applied to different disciplines of
science beyond the traditional research lines of physics, ranging from
biology, to economics and the social sciences
\cite{castellano2007sps}.  In particular, there has been a fruitful
effort in providing models of language dynamics, including dynamics of
language competition \cite{Abrams_2003}, language evolution
\cite{NowakKrak1999,Nowak_Komarova_2001} and semiotic dynamics
\cite{Steels1996,Baronchelli_JStatMech_2006}.

In the field of language competition, the Abrams-Strogatz model
\cite{Abrams_2003} has triggered the development of several models
which take into account the competition of many
\cite{Stauffer_2005,Tesileanu_2006} or few
\cite{Kosmidis_2005,Stauffer_Castelló_2006} languages. A review of
some of these models can be found in \cite{Schulze_2006_CiSE}.
Building up upon a proposal by Minett and Wang \cite{Wang_2005_TRENDS_Ecology,Minett_2008}, the
AB-model is a model of two non-excluding options, in which agents can
be in two symmetric states (A or B) and in a third mixed (AB) state of
coexisting options at the individual level \cite{Castello2006}. It has
been used to study the competition between two socially equivalent
languages, where AB agents are associated to bilingualism. The model
has been studied in two-dimensional and small world networks
\cite{Castello2006} and in networks with community structure
\cite{Castello2007,Toivonen_2008}. The final state of the system is
always consensus in one of the options, A or B.



In semiotic dynamics, the Naming
Game~\cite{Steels1996,Baronchelli_JStatMech_2006,Baronchelli_ng_long}
describes a population of agents playing pairwise interactions in
order to {\em negotiate} conventions, i.e., associations between forms
and meanings, and elucidates the mechanisms leading to the emergence
of a global consensus among them.  For the sake of simplicity the
model does not take into account the possibility of homonymy, so that
all meanings are independent and one can work with only one of them,
without loss of generality. An example of such a game is that of a
population that has to reach the consensus on the name (i.e. the form)
to assign to an object (i.e. the meaning) exploiting only local
interactions. However it is clear that the model, originally inspired
to robotic experiments \cite{Steels1996}, is appropriate to address
general situations in which negotiation rules a decision process on a
set of conventions (i.e. opinion dynamics, etc.). The Naming Game has been studied
in fully connected graphs (i.e. in mean-field or homogeneous mixing populations) \cite{Steels1996,Baronchelli_JStatMech_2006,Baronchelli_ng_long}, 
regular lattices \cite{Baronchelli_2005}, small world networks \cite{dallasta06}
and complex networks \cite{dallasta06b,dallasta06c}.
The final state of
the system is always consensus, but stable polarized states can be
reached introducing a simple confidence/trust
parameter \cite{Baronchelli_2007}.  In this paper, we shall focus on
the particular case in which only two options compete within the
population \cite{Baronchelli_ng_long,Baronchelli_2007}.

This paper is structured as follows: in Section~\ref{sec:2} we present
the microscopic description of the two models studied in this paper,
the Naming Game restricted to two conventions and the AB-model. In
Section~\ref{sec:3} we look at the macroscopic description of the
models, while in Sections~\ref{sec:4} and \ref{sec:5} we explore in
detail the differences between the two models arising from the
different microscopic interaction rules. Finally, we present in
Section~\ref{sec:6} the conclusions as well as a discussion about the
implications for language competition of the results obtained in this
paper.

\section{The models}
\label{sec:2}
We present here the two models considered in this paper: the
generalized Naming Game restricted to two conventions
\cite{Baronchelli_JStatMech_2006,Baronchelli_2007} and the AB-model
\cite{Castello2006}, extended in such a way that confidence/trust is
considered. In both models, we consider a set of $N$ interacting
agents embedded in a network. At each time step, and starting from a
given initial condition, we select randomly an agent and we update its
state according to the dynamical rules corresponding to each model.

In the Naming Game
\cite{Baronchelli_JStatMech_2006,Baronchelli_ng_long}, an agent is
endowed with an internal inventory in which it can store an a priori
unlimited number of conventions.  Initially, all inventories are
empty. At each time step, a pair of neighboring agents is chosen
randomly, one playing as ``speaker'', the other as ``hearer'', and
negotiate according to the following rules:

\begin{itemize} 
\item the speaker selects randomly one of its conventions and conveys
  it to the hearer (if the inventory is empty, a new convention is
  invented by the speaker);
\item if the hearer's inventory contains such a convention, the two
  agents update their inventories so as to keep only the convention
  involved in the interaction ({\em success});
\item otherwise, the hearer adds the convention to those already stored in its
inventory ({\em failure}).  
\end{itemize}

Here we are interested in the particular case in which a population
deals with only two competing conventions (say $A$ or $B$)
\cite{Baronchelli_ng_long}. We therefore assign to each agent one of
the two conventions at the beginning of the process, preventing in
this way further invention (that can happen only when the speaker's
inventory is empty). Moreover, we adopt the generalized Naming Game
scheme \cite{Baronchelli_2007}, in which a confidence/trust parameter
$\beta$ determines the update rule following a success: with
probability $\beta$ the usual dynamics takes place, while with the
complementary probability $1-\beta$ nothing happens. The usual case is
thus recovered for $\beta=1$. For brevity we shall refer to this
setting (generalized Naming Game restricted to two conventions) as the
2c-Naming Game. In this simplified case, it is easy to see that the
transition probabilities are the following~\cite{Baronchelli_2007}:
\begin{eqnarray}
  p_{A\to AB} = n_{B} + \frac{1}{2} n_{AB}, & & \; p_{B\to AB} =
  n_A + \frac{1}{2} n_{AB} \label{eq:NGa} \\
  p_{AB \to A} = \frac{3 \beta}{2} n_A + \beta n_{AB}, && p_{AB \to B} = \frac{3 \beta}{2} n_B + \beta n_{AB} \label{eq:NGb}
\end{eqnarray}

\noindent where $n_j$ ({\it j}=A, B, AB) are the fraction of agents
storing in their inventory the conventions $A$, $B$ or both $A$ and
$B$, respectively.


For the AB-model \cite{Castello2006}, an agent can be in three
possible states: {\it A}, choosing option A (using language A), {\it
  B}, choosing option B (using language B) or {\it AB}, choosing both,
A and B (using both languages, {\it bilingual} agent). An agent
changes its state with a probability which depends on the fraction of
agents in the other states, $n_j$ ({\it j}=A, B, AB). The transition
probabilities are the following\footnote{In the original model
  \cite{Wang_2005_TRENDS_Ecology} languages with different prestige were
  considered. The factor $1/2$ in the AB-model corresponds to the case
  of socially equivalent languages.}:
\begin{eqnarray}
p_{A\to AB} = \frac{1}{2} n_B, & & \; p_{B\to AB} =
\frac{1}{2} n_A \label{eq:ABa} \\
p_{AB \to A} = \frac{1}{2} \beta (1-n_B), && p_{AB \to B} = \frac{1}{2} \beta ( 1- n_A) \label{eq:ABb}
\end{eqnarray}

An agent changes from the A or B state towards the AB state
(equation~(\ref{eq:ABa})), with a probability proportional to the
agents in the opposite option. The probability that an AB agent moves
towards the A or B state (equation~(\ref{eq:ABb})) is proportional to
the density of agents sharing that option, including those in the AB
state ($1-n_i=n_j+n_{AB}$; $i,j=A,B$, $i\neq j$). In this paper, we
extend the original model in analogy to the extension proposed for the
Naming Game in \cite{Baronchelli_2007}: an agent abandons an option or
language according to the dynamics of the AB-model (changes from {\it
  AB} to {\it A} or {\it B}) with a probability $\beta$, while with a
probability $1- \beta$ nothing happens. In the context of language
competition, the parameter $\beta$ can be interpreted as an inertia to
stop using a language, and at the same time, as a reinforcement of the
status of being bilingual, which was not taken into account in the
original model (recovered by setting $\beta=1$).

In both models, a unit of time is defined as $N$ iterations, so that
at every unit of time each agent has been updated on average once. To
describe the dynamics of the system we use as an order parameter the
\emph{interface density} $\rho$, defined as the fraction of links
connecting nodes in different states. When the system approaches
consensus, domains grow in size, and the interface density
decreases. Zero interface density indicates that an absorbing state,
consensus, has been reached. We also use the average interface
density, $\langle \rho \rangle$, where $\langle \cdot \rangle$
indicates average over realizations of the stochastic dynamics
starting from different random initial conditions.

\section{Macroscopic description}
\label{sec:3}

In Section~\ref{sec:2} we have presented the microscopic description
of the 2c-Naming Game and the AB-model, i.e., the set of local
interactions among the agents.  In order to have a macroscopic
description of the dynamical evolution of the system as a whole we
derive the mean-field equations for the fraction of agents in each
state. For the 2c-Naming Game one has~\cite{Baronchelli_2007}:
\begin{eqnarray}
\frac{d n_A}{dt} = - n_A n_B + \beta n_{AB}^{2} + \frac{3 \beta -1}{2}n_A n_{AB} 
\label{eq:meanfield_NG_a} \\
\frac{d n_B}{dt} = - n_A n_B + \beta n_{AB}^{2} + \frac{3 \beta -1}{2}n_B n_{AB}
\label{eq:meanfield_NG_b}
\end{eqnarray}

\noindent and $n_{AB}=1-n_A-n_B$. 

The stability analysis showed that
there exist three fixed points \cite{Baronchelli_2007}: (1)
$n_A=1,n_B=0,n_{AB}=0$; (2) $n_A=0,n_B=1,n_{AB}=0$ and (3)
$n_A=b(\beta),n_B=b(\beta),n_{AB}=1-2b(\beta)$ with
$b(\beta)=\frac{1+5\beta-\sqrt{1+10\beta+17\beta^2}}{4\beta}$ (and
b(0)=0). A non-equilibrium phase transition occurs for a critical
value $\beta_c=1/3$. For $\beta_c>1/3$ consensus is stable. For
$\beta_c<1/3$ a change of stability gives place to a stationary
coexistence of $n_A=n_B$ and a finite density of undecided agents
$n_{AB}$, fluctuating around the average values $b(\beta)$ and
$1-2b(\beta)$. In the Naming Game with invention, in fact, the one
observed at $\beta_c=1/3$ is the first of a series of transitions
yielding the asymptotic survival of a diverging (in the thermodynamic
limit) number of conventions as $\beta \rightarrow 0$
\cite{Baronchelli_2007}.

For the AB-model one has:
\begin{eqnarray}
\frac{d n_A}{dt} = \frac{1}{2}(- n_A n_B + \beta n_{AB}^{2} + \beta n_A n_{AB}) 
\label{eq:meanfield_AB_a} \\
\frac{d n_B}{dt} = \frac{1}{2}(- n_A n_B + \beta n_{AB}^{2} + \beta n_B n_{AB})
\label{eq:meanfield_AB_b}
\end{eqnarray}

\noindent and $n_{AB}=1-n_A-n_B$. 

The stability analysis shows that there exist three fixed points: (1)
$n_A=1,n_B=0,n_{AB}=0$; (2) $n_A=0,n_B=1,n_{AB}=0$ and (3)
$n_A=f(\beta),n_B=f(\beta),n_{AB}=1-2f(\beta)$ with
$f(\beta)=\frac{3\beta-\sqrt{\beta(\beta+4)}}{2(2\beta-1)}$.

Notice that in both models, at $\beta=0$ the third fixed point becomes
a stable absorbing state in which the system reaches consensus in the
AB-state.


Surprisingly, the two original models ($\beta=1$) are equivalent in
the mean-field approximation. There is just a different time scale
coming from the prefactor $1/2$ in the AB-model
(equations~(\ref{eq:meanfield_AB_a}) and
(\ref{eq:meanfield_AB_b})). The mean-field approximation is exact in
the thermodynamic limit, and valid for large systems in fully
connected networks.
However, the two models differ at their local interactions
(see equations (\ref{eq:NGa}-\ref{eq:ABb}) for $\beta=1$). To explore
the effects of these differences at the microscopic level, in
Section~\ref{sec:4} we investigate, in a complete graph, the role of 
the parameter $\beta$ as described in equations (\ref{eq:meanfield_NG_a}-\ref{eq:meanfield_AB_b}), 
while in Section~\ref{sec:5} we focus on the interface motion in regular lattices for the
original case $\beta=1$.

\section{Phase transition}
\label{sec:4}

\begin{figure}
\vspace{0.5cm}
\includegraphics[angle=0,scale=0.3]{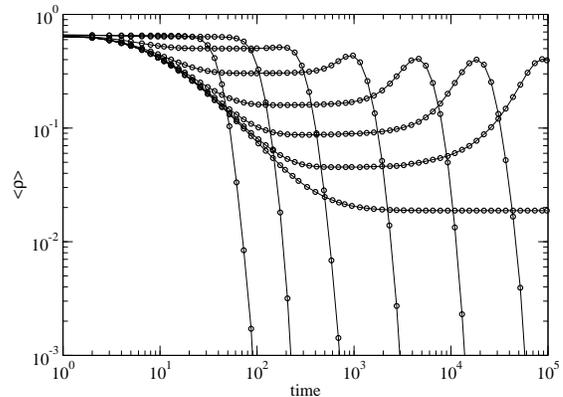}
\caption{AB-model: time evolution of the average interface density,
  $\langle \rho \rangle$, in a fully connected network of $N=10000$
  agents for different values of $\beta$. From left to right:
  $\beta=1.0, 0.2, 0.05, 0.01, 0.002, 0.0005, 0.0001, 0.0$. Averaged
  over 1000 runs.}
\label{fig:1}       
\end{figure}
\begin{figure}
\vspace{0.5cm}

  \includegraphics[angle=0,scale=0.3]{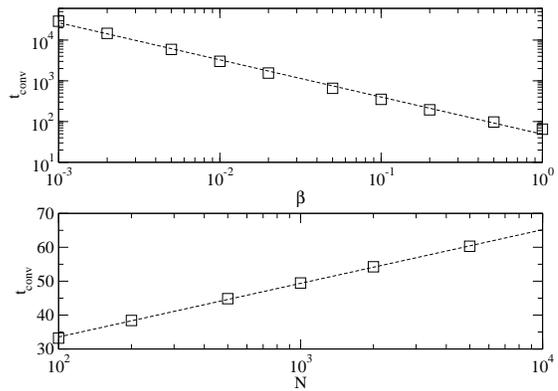}
  \caption{AB-model. Top: scaling of the time of convergence with
    $\beta$ for $N=10000$: $t_{conv} \sim \beta^{-1}$. Averaged over
    200-800 runs depending on the value of $\beta$. Bottom: scaling of
    the time of convergence with system size $N$ for $\beta=1$:
    $t_{conv} \sim ln(N)$. Averaged over 10000 runs.}
\label{fig:2}       
\end{figure}

Here we consider the extension of the AB-model presented above in a
fully connected network, with the aim to explore a possible
non-equilibrium phase transition in $\beta$ similar to the one found
in the 2c-Naming Game. In Figure~\ref{fig:1} we show the time
evolution of the average interface density, $\langle \rho \rangle$,
for different values of the parameter $\beta$. For large values of
$\beta$, $\langle \rho \rangle$ reaches a plateau followed by a finite
size fluctuation that drives the system to an absorbing
state. However, for $\beta\lesssim0.01$ we observe that after $\langle
\rho \rangle$ reaches the plateau, it increases again, reaching a
maximum value after which a finite size fluctuation drives the system
to consensus. 
In finite systems and for $\beta=0$, the system reaches a constant value of
$\langle \rho \rangle$, 
a frozen state corresponding to almost consensus in the AB-state, except for a small fraction of agents (less than 1\% on average for $N=10000$). This fraction decreases as $N$ increases, and complete consensus in the AB-state is reached in the thermodynamic limit (see stability analysis in the previous section). For $\beta=0$, $p_{AB \to A} =p_{AB \to B} =0$, so the
only possible evolution is that A and B agents move towards the AB
state. At the last stage of the dynamics, when $n_A$ and $n_B$ approach to zero, as soon as one of the two single-option densities, $n_i$, vanishes, $n_j$ remains constant $(i,j=A,B, i\neq j)$ because $p_{j\to AB} \sim n_i$, giving rise to the small fraction of agents in the state $j$ present in the frozen state.
The time to convergence scales with beta as $t_{conv} \sim
\beta^{-1}$ (Figure~\ref{fig:2}-top), as observed for the
2c-Naming Game for $\beta>\beta_c$ ($t_{conv} \sim
(\beta-\beta_c)^{-1}$) \cite{Baronchelli_2007}.

Contrary to the phase transition described in Section~\ref{sec:3}
obtained in the 2c-Naming Game \cite{Baronchelli_2007}, there is no
phase transition in the AB-model: at $\beta=0$, the system reaches
trivially a frozen state (dominance of the AB-state, with complete consensus in the thermodynamic limit); while for $\beta>0$ the final
absorbing state is, as usual, consensus in the A or B option.
Even though the two original models are equivalent in the mean-field
approximation (case $\beta=1$), we observe two different
behaviors when the parameter $\beta$ is taken into account. 
This can be shown formally by looking at the time evolution
of the {\it magnetization}, $m\equiv n_A-n_B$. For the 2c-Naming Game
and the AB-model, we have respectively:
\begin{eqnarray}
  \frac{d m}{dt} = \frac{3\beta-1}{2}n_{AB}m
  \label{eq:magnetization_NG} \\
  \frac{d m}{dt} = \frac{1}{2}\beta n_{AB}m
\label{eq:magnetization_AB}
\end{eqnarray}
In equation~(\ref{eq:magnetization_NG}), we can observe the origin of
the non-equilibrium phase transition described in Section~\ref{sec:3}
for the 2c-Naming Game. The time derivative of the magnetization,
$\frac{d m}{dt}$, vanishes at the critical point $\beta_c=1/3$.  For
$\beta_c>1/3$, $sign(\frac{d m}{dt})=sign(m)$, and therefore
$|m|\rightarrow 1$: the system is driven to an absorbing state of
consensus in the A or B option. For $\beta_c<1/3$, $sign(\frac{d
  m}{dt})=-sign(m)$ and $|m|\rightarrow 0$, giving rise to stationary
coexistence of the three phases, with $n_A=n_B$ and a finite density
of AB agents.
For the AB-model, instead, we can see in
equation~(\ref{eq:magnetization_AB}) that for $\beta > 0$,
$sign(\frac{d m}{dt})=sign(m)$ so that consensus in the A or B option is always reached. The
time derivative of the magnetization, $\frac{d m}{dt}$, vanishes at
$\beta=0$, where the dynamics gets stuck in an absorbing state corresponding to consensus in the AB state in the thermodynamic limit. Therefore, the phase
transition observed in the 2c-Naming Game is not observed.

The reason of the difference shown above has to be found in the
differences that these models have at the microscopic level. The fact
that in the AB-model A and B agents do not feel the influence of AB
agents is the key point which explains the different nature of the
transition for this model. For the case $\beta=1$, in the 2c-Naming
Game the second term in equation~(\ref{eq:NGa}) (influence of AB
agents on the A or B agents) and the first term in
equation~(\ref{eq:NGb}) (influence of the A or B agents on the AB
agents) combine in such a way that the mean-field equations are
equivalent to the ones in the AB-model. However, when $\beta < 1$
the combination of these terms originate the phase transition from an
absorbing final state towards a dynamical stationary state of
coexistence, as found in \cite{Baronchelli_2007}.

In Figure~\ref{fig:2}-bottom, we observe for the AB-model the
scaling of $t_{conv}$ with the system size for $\beta=1$: $t_{conv}
\sim ln(N)$, indicating that $t_{conv}$ increases slowly with the
system size.  As expected, this compares properly to the same scaling
already obtained for the 2c-Naming Game in \cite{Baronchelli_ng_long},
since the two models are equivalent in the mean field for $\beta=1$.

\begin{figure}
\vspace{0.5cm}
\includegraphics[angle=0,scale=0.3]{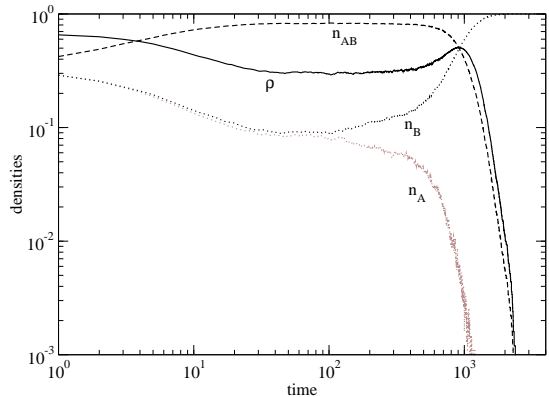}
\caption{AB-model. Density of agents in state A, $n_A$ (dotted gray), in state
  B, $n_B$ (dotted black), in state AB, $n_{AB}$ (dashed); and
  interface density, $\rho$ (solid line) for a typical realization of
  the AB-model. In the plateau, $n_A \sim n_B \simeq 0.1$, while the
  majority of agents are in the AB-state. Fully connected network,
  $\beta=0.01$ and $N=10000$ agents.}
\label{fig:3}       
\end{figure}

To understand the time evolution of the average interface
density $\langle \rho \rangle$ shown in Figure~\ref{fig:1} for
$\beta\lesssim0.01$, we show in Figure~\ref{fig:3} the time evolution
of $\rho$ and the densities of agents in each state, $n_A$, $n_B$ and
$n_{AB}$, for a typical realization of the dynamics and a given small
$\beta$. Because of the inertia of the AB agents to move away from
their state (small $\beta$), at the beginning we observe an 
increase of $n_{AB}$ together with the corresponding
decrease of $n_A$, $n_B$ and $\rho$. Then, $\rho$ and the 
three densities reach a plateau. Most of the
agents are in the AB state, while a competition between options A and
B takes place, with $n_A \simeq n_B < n_{AB}$. This metastable state
lasts longer as we increase the system size. At a certain point, however, 
a system size fluctuation drives the density of one of the two states 
(A in the figure) to zero, while the other (B in the figure) starts gaining 
ground. Since there are less and less agents in the state becoming extinct, and
agents having one option do feel only the presence of agents in the
opposite state, agents in the dominant state become more and more
stable, until, when the other state disappears, they become totally 
stable. During this process, the interface density increases as
$n_{AB}$ decreases. The peak of $\rho$ corresponds to the point where
$n_{AB}=n_i$ (being {\it i} the state which takes over the whole system,
B in the figure). When one of the states has vanished (A in the
figure), the AB agents slowly move towards the remaining state 
(B in the figure) and the system reaches consensus.

\section{Interface dynamics: 1-d and 2-d lattices}
\label{sec:5}

We study and compare here the interface dynamics in regular lattices
with periodic boundary conditions for the two original models
($\beta=1$).  The 2c-Naming Game has been shown to exhibit a diffusive
interface motion in a one-dimensional lattice, with a diffusion
coefficient $D=401/1816 \simeq 0.221$ \cite{Baronchelli_2005}. We therefore 
focus on the AB-model, and, to analyze the interface dynamics in a
one-dimensional lattice with $N$ agents, we consider a single
interface between two linear clusters of agents. In each of the
clusters, all the agents are in the same state. We consider a cluster
of agents in the state A on the-left and another cluster of agents in
the B state on the right. We call $C_m$ an interface of $m$ agents in
state C (for clarity, here C labels an AB agent). Due to the dynamical
rules, the only two possible interface widths are $C_0$, corresponding
to a two directly connected clusters $\cdot\cdot AAABBB \cdot\cdot$,
or $C_1$, corresponding to an interface of width one, $\cdot\cdot
AAACBBB \cdot\cdot$.  It is straightforward to compute the probability
$p_{0,1}=1/2N$ that a $C_0$ interface becomes a $C_1$ in a single time
step. Otherwise, it stays in $C_0$. In the same way,
$p_{1,0}=1/2N$. We are now interested in determining the stationary
probabilities of the Markov chain defined by the transition matrix

\begin{equation}
 M= \left(
\begin{array}{cc}
  1-\frac{1}{2N} & \frac{1}{2N} \\
  \frac{1}{2N} & 1-\frac{1}{2N} 
\label{eq:Markov_matrix}
\end{array}
\right)
\end{equation}

\noindent in which the basis is $\{C_0,C_1\}$. The stationary
probability vector, ${\bf P}=\{P_0, P_1\}$ is computed by imposing
${\bf P}(t+1)-{\bf P}(t)=0$, i.e., $(M^T-I){\bf P}=0$. We obtain
$P_0=1/2, P_1=1/2$.  Since the interface has a bounded width, we
assume that it can be modeled as a point-like object localized at
position $x=(x_l+x_r)/2$, where $x_l$ is the position of the rightmost
site of cluster A, and $x_r$ the leftmost site of cluster B. An
interaction $C_m \rightarrow C_{m^{'}}$ corresponds to a set of
possible movements for the central position $x$. We denote by $W(x
\rightarrow x\pm\delta)$ the transition probability that an interface
centered in $x$ moves to to the position $x\pm\delta$. The only
possible transitions are: $W(x \rightarrow
x\pm\frac{1}{2})=\frac{1}{4N}P_0+\frac{1}{4N}P_1$. Using the results
obtained for the stationary probability vector we get $W(x \rightarrow
x\pm\frac{1}{2})=\frac{1}{4N}$.
We are now able to write the master equation for the probability ${\it
  P}(x,t)$ to find the interface in position $x$ at time $t$. In the
limit of continuous time and space:

\begin{figure}
\vspace{0.5cm}
\includegraphics[angle=0,scale=0.3]{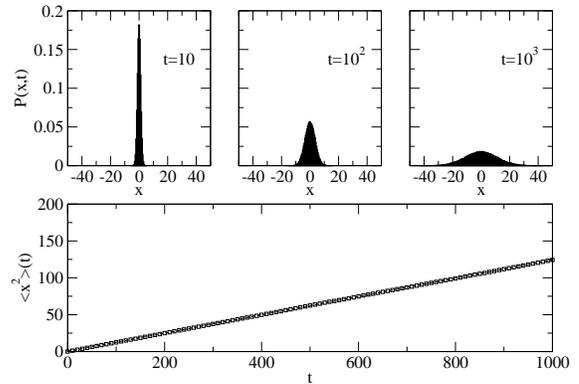}
\caption{AB-model: evolution of the position of an interface in a
  one-dimensional regular lattice. {\it Top}: time evolution of the
  distribution $P(x,t)$. {\it Bottom}: time evolution of the mean
  square displacement $\langle x^2(t) \rangle=2D_{exp}t$.
  The value $D_{exp}=0.06205$ obtained from the fitting is in perfect 
  agreement with the theoretical prediction $D=1/16=0.0625$.}
\label{fig:4}       
\end{figure}

\begin{eqnarray} 
{\it P}(x,t+1)-{\it P}(x,t) \approx \delta t
  \frac{\partial {\it P}(x,t)}{\partial t}
  \label{eq:continous_time}, \\
  {\it P}(x+\delta x,t) \approx 
  {\it P}(x,t) +\delta x \frac{\partial {\it P}(x,t)}{\partial x} +\\ 
  \nonumber + \frac{1}{2}(\delta x)^2 \frac{\partial^2 {\it P}(x,t)}{\partial x^2}
\label{eq:continous_space}
\end{eqnarray}

\noindent In this limit, the master equation reads:
\begin{eqnarray}
  \frac{\partial {\it P}(x,t)}{\partial t} = \frac{D}{N} \frac{\partial^2 {\it P}(x,t)}{\partial x^2}
\label{eq:diffusion_equation}
\end{eqnarray}

\noindent where the diffusion coefficient is $D=1/16=0.0625$ (in the
appropriate dimensional units $(\delta x)^2/\delta t$).  These
analytical results are confirmed by numerical simulations. In
Figure~\ref{fig:4} we show the time evolution of ${\it P}(x,t)$, which
displays a clear diffusive behavior. The mean-square distance follows
a diffusion law $\langle x^2 \rangle=2D_{exp}t$, with $D_{exp}=0.06205$
being the diffusion coefficient obtained numerically.

\begin{figure}
\centering
\includegraphics[angle=0,scale=0.31]{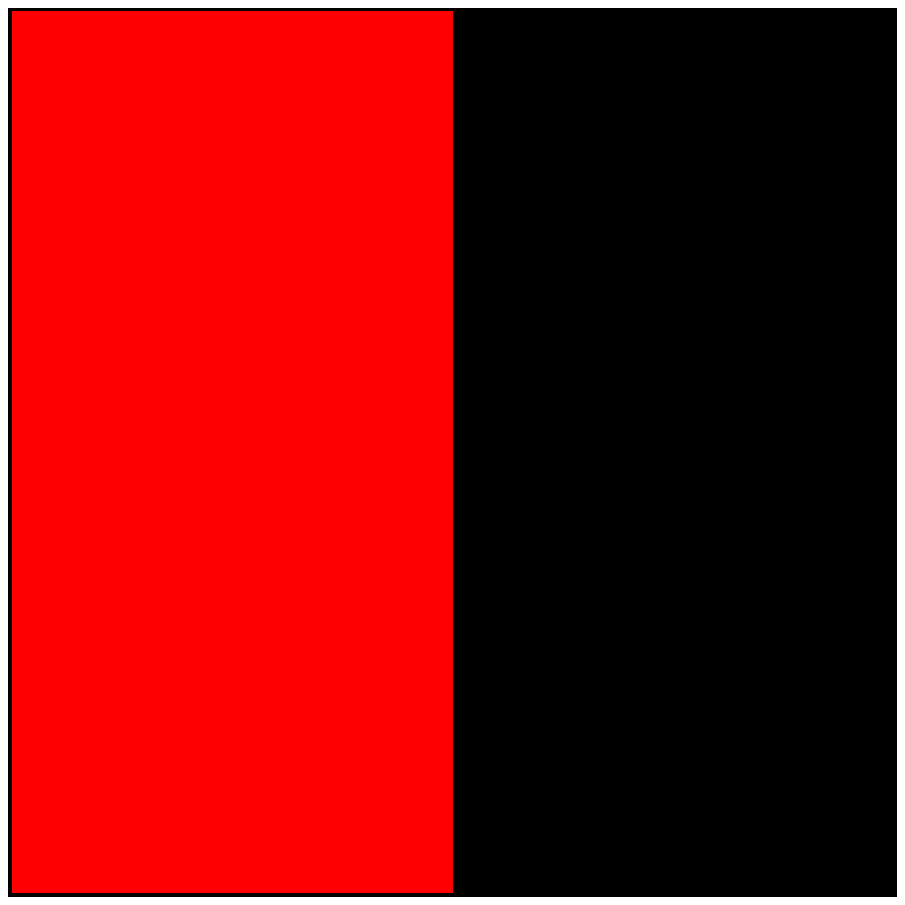}
\includegraphics[angle=0,scale=0.31]{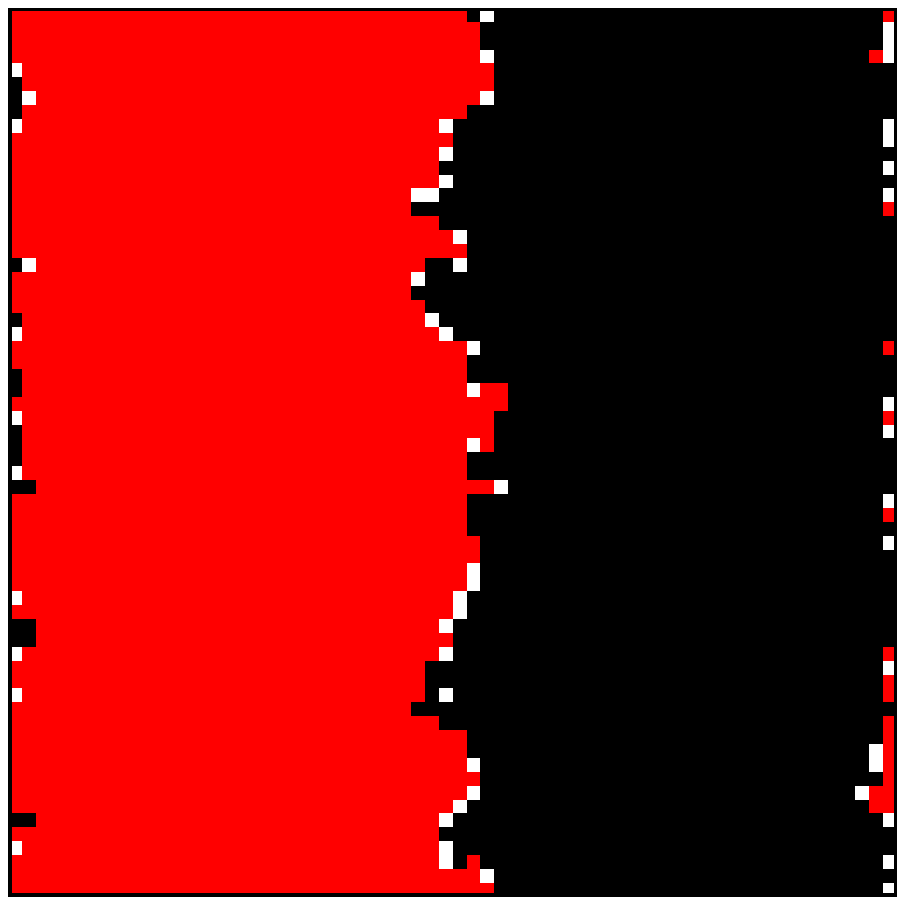}
\includegraphics[angle=0,scale=0.31]{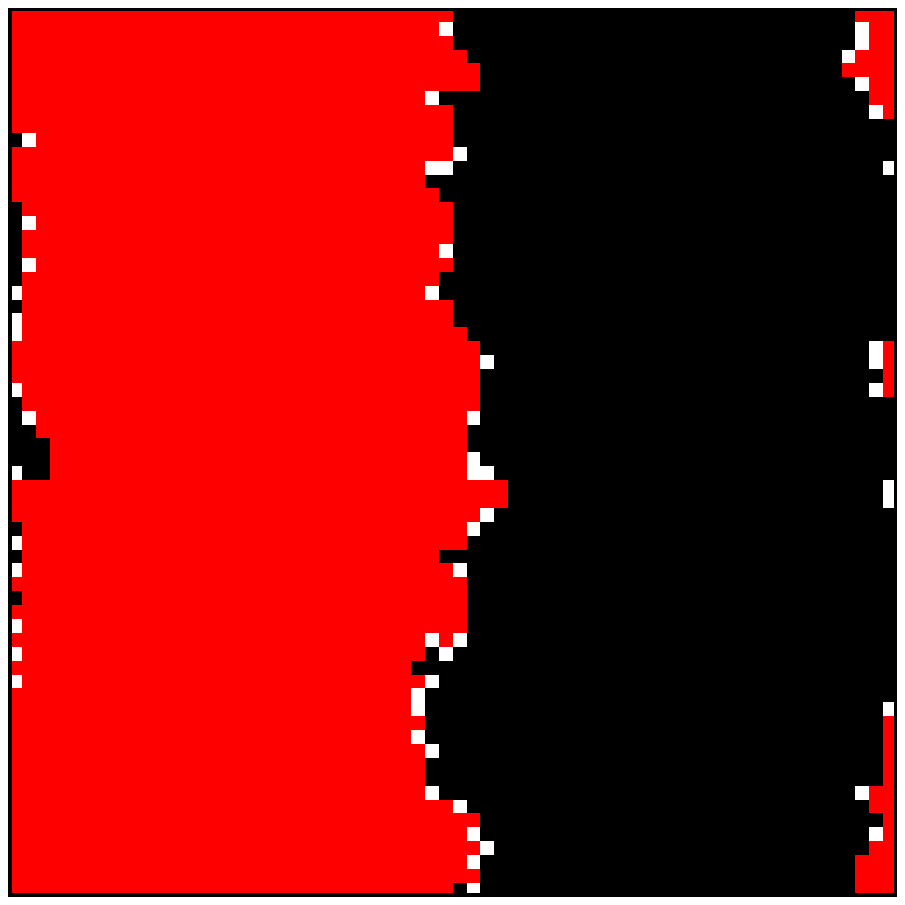}

\vspace{0.75cm}

\includegraphics[angle=0,scale=0.31]{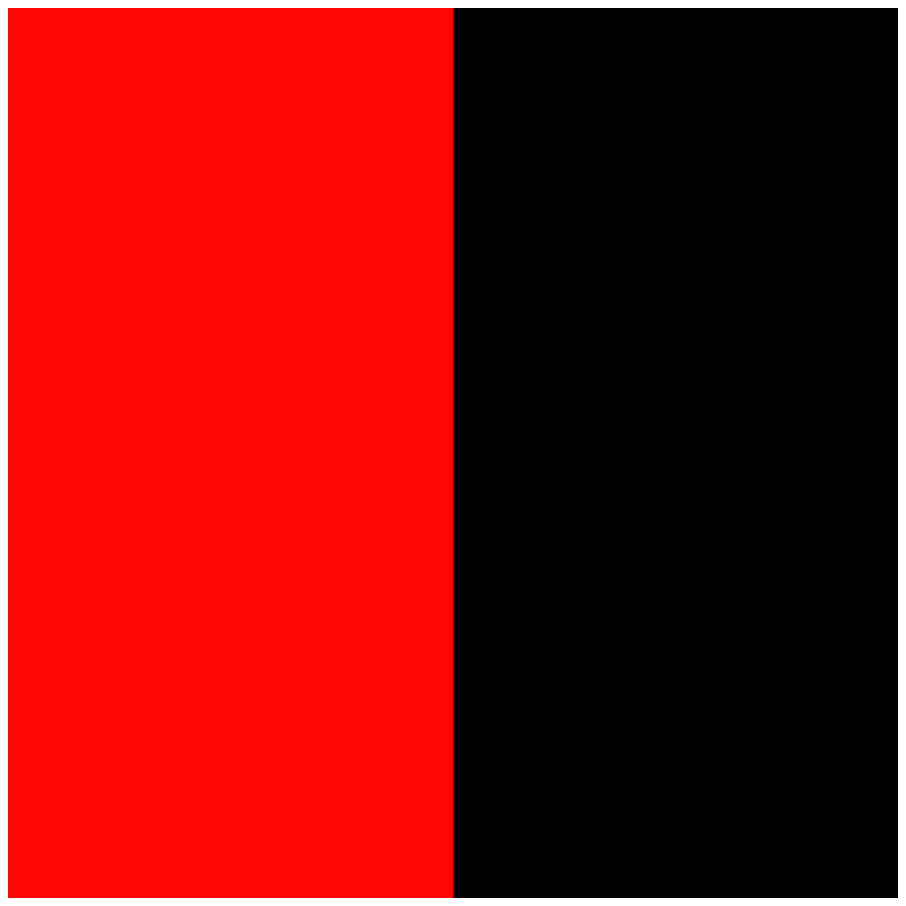}
\includegraphics[angle=0,scale=0.31]{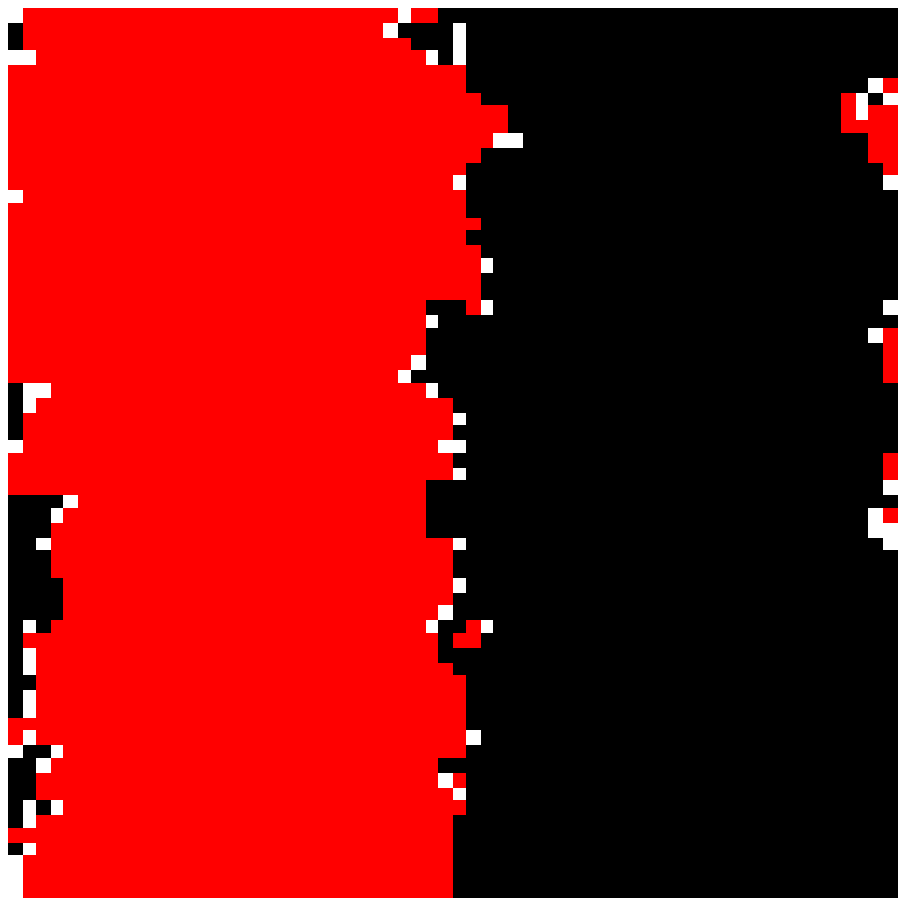}
\includegraphics[angle=0,scale=0.31]{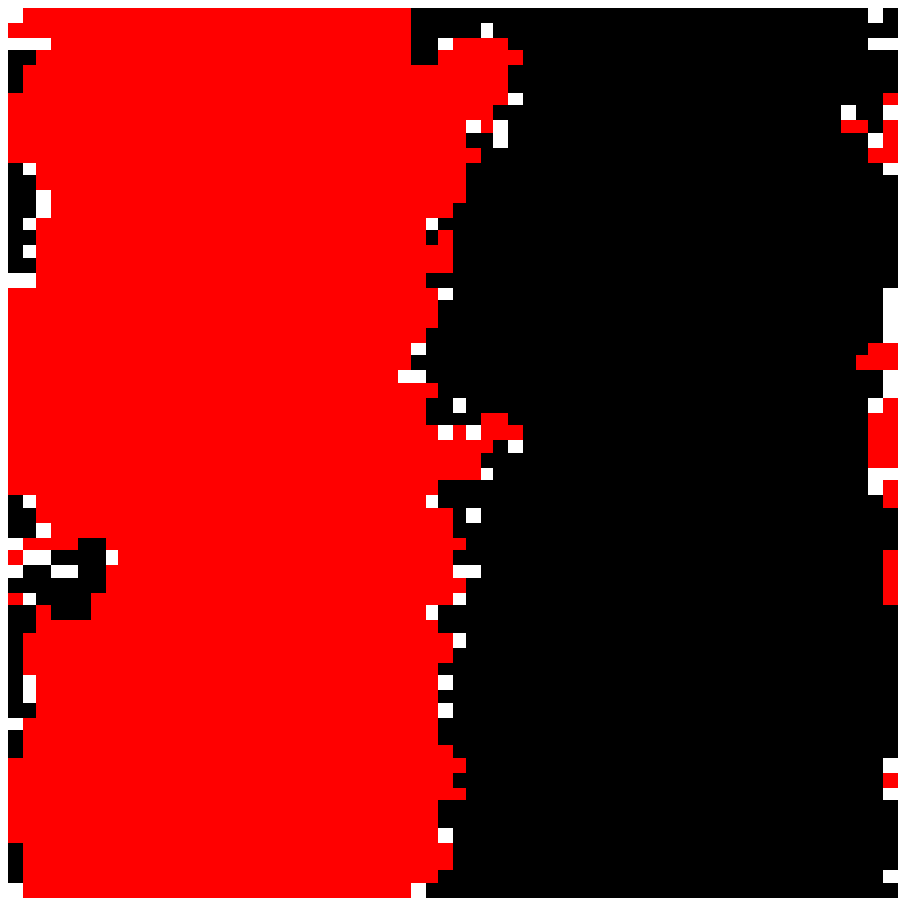}

\vspace{20pt}

%
\caption{Initial conditions with one half of the lattice in the A
  state, and the other half in the B state. $N=64^2$ {\it Top}:
  AB-model. Stripe-like metastable state. t=0, 100, 200 from left to
  right. {\it Bottom}: 2c-Naming Game: t=0, 50, 100 from left to
  right. Snapshots are selected taking into account the different time scale coming from the prefactor $1/2$ in the meanfield equations (7) and (8).}
\label{fig:5}       
\end{figure}

Thus, the AB-model and the
2c-Naming Game display the same diffusive interface motion in
one-dimensional lattices, but they differ in about one order of magnitude
in the diffusion coefficient, indicating that in the AB-model
interfaces diffuse much slower. It can also be seen that the growth of
the typical size of the clusters is $\zeta\sim t^{\alpha}$, with
$\alpha\simeq 0.5$, leading to the well known coarsening process found
also in SFKI models \cite{Gunton_1983} (not shown).

In two-dimensional lattices, on the other hand, it has been
shown that starting from random initially distributed options
among the agents, both models present a coarsening $\zeta\sim
t^{\alpha}$, $\alpha\simeq 0.5$, with a curvature driven interface dynamics \cite{Castello2006,Baronchelli_2005} and AB-agents placing
themselves at the interfaces between single-option domains.
In Figure~\ref{fig:5} we show snapshots comparing the two dynamics,
starting from initial conditions where we have half of the lattice in
state A, and the other half in state B. Given that the interface
dynamics is curvature driven, flat boundaries are very stable. In both
models these stripe-like configurations give rise to metastable
states, already found in \cite{Castello2006} for the AB-model:
dynamical evolution of boundaries close to flat interfaces but with
interfacial noise present. These configurations evolve by diffusion of
the two walls (average interface density fluctuating around a fixed
value) until they meet and the system is driven to an absorbing
state. In the AB-model, also when starting from options randomly
distributed through the lattice, $1/3$ of the realizations end up in
such stripe-like metastable states \cite{Castello2006}.We checked that the 
same turns out to be true also for the 2c-Naming Game. In the usual Naming Game 
with invention, on the other hand, stripes are better avoided since in 
that case the two convention state is usually reached when one cluster is 
already considerably larger than the other.

\begin{figure}
\vspace{0.5cm}
\includegraphics[angle=0,scale=0.3]{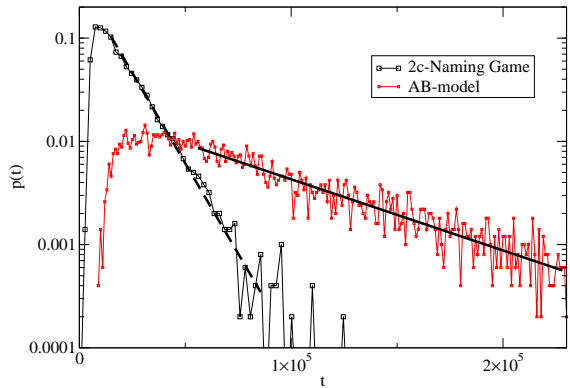}
\caption{Probability distribution for the time to reach consensus,
  starting with stripe-like configurations. Black: 2c-Naming Game,
  $\tau_{NG}\simeq 1.2\e{4}$; gray: AB-model, $\tau_{AB}\simeq
  6.3\e{4}$. Averages are over 5000 runs.}
\label{fig:6}       
\end{figure}

We show in Figure~\ref{fig:6} the distribution of survival times for
the two models, i.e., the time needed for a stripe-like configuration
to reach an absorbing state. The distribution displays an exponential
tail, $p(t)\sim e^{-t/\tau}$ with a characteristic time $\tau$. The
characteristic time for the AB-model is however
larger than the one for the 2c-Naming Game ($\tau_{AB}>\tau_{NG}$),
confirming that the AB-model interface dynamics slows down the diffusion 
of configurations such as stripes in two dimensional lattices, or walls in one
dimensional lattices. Notice that in both cases, the differences found are beyond the trivial different time scale corresponding to the prefactor $1/2$ in the mean field equations for the AB-model (equations (7) and (8)).

\section{Discussion and conclusion}
\label{sec:6}

We have analyzed and compared the 2c-Naming Game and the AB-model, 
originally defined in the context of language emergence and competition,
respectively. We have shown that although these two models are
equivalent in mean-field, their microscopic differences give rise to
different behaviors. In particular, we have focused on
(1) the extension of the models by introducing the parameter $\beta$,
describing the inertia of the agents to abandon an acquired option,
and (2) the interface dynamics in one and two-dimensional lattices.

As for the extension of the models incorporating the parameter
$\beta$, even though the original models are equivalent in the mean
field approximation for $\beta=1$, an important difference concerns
the existence of a phase transition. While the 2c-Naming Game features
a non-equilibrium phase transition between consensus and stationary
coexistence of the three phases present in the system, in the
AB-model such a transition does not exist, the model featuring a
trivial frozen state for $\beta=0$ (dominance of the AB-state, with complete consensus in the thermodynamic limit), and the usual
consensus in the A or B state, for $0< \beta \leq 1$.

As for the interface dynamics, we have shown in one-dimensional
lattices that the AB-model has a diffusive interface motion analogous
to the one already found in the 2c-Naming Game, but with a diffusion
coefficient nearly one order of magnitude smaller. In two-dimensional
lattices, we have studied the time evolution of stripe-like
configurations, which are metastable in both models but have a larger
life time in the AB-model. Both results indicate that in comparison to
the 2c-Naming Game, the AB-model interface dynamics slows down the
diffusion of these configurations.

It is interesting to discuss the implications of our results on the AB-model in
the context of dynamics of language competition. In the original
AB-model ($\beta=1$), the density of bilingual individuals remains
small during the language competition process (around 20\%), and in the 
end bilingual individuals disappear together with the language facing extinction.
When introducing the parameter $\beta$, interpreted here as a sort of
inertia to stop using a language, and, at the same time, as a
reinforcement of the status of being bilingual, we observed the
following. When $\beta$ is small enough, bilingual agents rapidly become the
majority, while the two monolingual communities compete between each
other and have similar sizes. The smaller the parameter $\beta$,
the larger the bilingual community is at this point. After this stage
of coexistence, a symmetry breaking takes place and one of the two
monolingual communities starts to grow, while the other looses
ground. When this monolingual community faces extinction, the language
spoken in that community survives in the bilingual community until
this community also disappears. 

In other words, contrary to the
original model, the extinction of a language takes place in two steps.
At first, the agents who speak just that language disappear, but
the language does not, as it is still spoken in the society by the
bilingual agents. Then bilingual agents disappear, too, 
which leads to the extinction of the language. Within the limited
framework of the AB-model, in which there does not exist any political
measure enhancing the prestige of an endangered language, these
results come to support the idea that, in societies with two languages,
the disappearance of a monolingual community using a language as its only way of
communication could represent the first step in the extinction of that language.
The other language could indeed become eventually the only spoken one, 
as the bilingual agents would eventually end up using only the language spoken 
by the remaining monolingual community.

The dynamics of the original Naming Game as well as that of the
AB-model are strongly affected by the underlying interaction 
network, as it has been shown
in~\cite{Baronchelli_2005,Castello2006,dallasta06,dallasta06b,dallasta06c,Castello2007,Toivonen_2008,Baronchelli_2007}
(times to consensus, apparition of trapped metastable states, etc). In
order to understand the implications of different complex social networks (small world
effect, community structure, etc) for the extension of the model
presented here, it is worth investigating in future this
extension in topologies of increasing complexity.


\section*{Acknowledgments}
\label{sec:7}

X. Castell\'o acknowledges financial support from a phD fellowship of
the Govern de les Illes Balears (Spain).  A. Baronchelli acknowledges
support from the DURSI, Generalitat de Catalunya (Spain) and from
Spanish MEC (FE\-DER) through project No: FIS2007-66485-C02-01. 
This research has been party supported by the TAGora project funded by the
Future and Emerging Technologies program (IST-FET) of the European
Commission under the European Union RD contract IST-034721.


\begin{thebibliography}{10}

\bibitem{castellano2007sps}
C.~Castellano, S.~Fortunato, and V.~Loreto.
\newblock Statistical physics of social dynamics.
\newblock {\em Reviews of Modern Physics}, 2007.
\newblock (in press) [e-print arXiv:0710.3256].

\bibitem{Abrams_2003}
D.~M. Abrams and S.~H. Strogatz.
\newblock Modelling the dynamics of language death.
\newblock {\em Nature}, 424:900, 2003.

\bibitem{NowakKrak1999}
M.A. Nowak and D.C. Krakauer.
\newblock The evolution of language.
\newblock {\em PNAS}, 96(14):8028--8033, 1999.

\bibitem{Nowak_Komarova_2001}
M.~A. Nowak, N.~L. Komarova, and P.~Niyogi.
\newblock Evolution of universal grammar.
\newblock {\em Science}, 291(5501):114--118, 2001.

\bibitem{Steels1996}
L.~Steels.
\newblock A self-organizing spatial vocabulary.
\newblock {\em Artificial Life}, 2(3):319--332, 1995.

\bibitem{Baronchelli_JStatMech_2006}
A.~Baronchelli, M.~Felici, V.~Loreto, E.~Caglioti, and L.~Steels.
\newblock Sharp transition towards shared vocabularies in multi-agent systems.
\newblock {\em Journal of Statistical Mechanics}, P06014, 2006.

\bibitem{Stauffer_2005}
D.~Stauffer and C.~Schulze.
\newblock Microscopic and macroscopic simulation of competition between
  languages.
\newblock {\em Physics of Life Reviews}, 2:89--116, 2005.

\bibitem{Tesileanu_2006}
T.~Tesileanu and H.~Meyer-Ortmanns.
\newblock Competition of languages and their hamming distance.
\newblock {\em International Journal of Modern Physics C}, 17:259--278, 2006.

\bibitem{Kosmidis_2005}
K.~Kosmidis, J.~M. Halley, and P.~Argyrakis.
\newblock Language evolution and population dynamics in a system of two
  interacting species.
\newblock {\em Physica A}, 353:595--612, 2005.

\bibitem{Stauffer_Castelló_2006}
D.~Stauffer, X.~Castell\'o, V.~M. Egu\'iluz, and M.~San~Miguel.
\newblock Microscopic Abrams-Strogatz model of language competition.
\newblock {\em Physica A}, 374:835--842, 2007.

\bibitem{Schulze_2006_CiSE}
C.~Schulze and D.~Stauffer.
\newblock Recent developments in computer simulations of language competition.
\newblock {\em Computing in Science and Engineering}, 8:60--67, 2006.

\bibitem{Wang_2005_TRENDS_Ecology}
William S-Y. Wang and James~W. Minett.
\newblock The invasion of language: emergence, change and death.
\newblock {\em Trends in Ecology and Evolution}, 20:263--269, 2005.

\bibitem{Minett_2008}
J.~W. Minett and W.~S.-Y. Wang.
\newblock Modelling endangered languages: the effects of bilingualism and
  social structure.
\newblock {\em Lingua}, 118:19--45, 2008.

\bibitem{Castello2006}
X.~Castell\'o, V.~M. Egu\'{\i}luz, and M.~San~Miguel.
\newblock Ordering dynamics with two non-excluding options: Bilingualism in
  language competition.
\newblock {\em New Journal of Physics}, 8:308--322, 2006.

\bibitem{Castello2007}
X.~Castell\'o, R.~Toivonen, V.~M. Egu\'{\i}luz, J.~Saram\"aki, K.~Kaski, and
  M.~San~Miguel.
\newblock Anomalous lifetime distributions and topological traps in ordering
  dynamics.
\newblock {\em Europhysics Letters}, 79:66006 (1--6), 2007.

\bibitem{Toivonen_2008}
R.~Toivonen, X.~Castell\'o, V.~M. Egu\'{\i}luz, J.~Saram\"aki, K.~Kaski, and
  M.~San~Miguel.
\newblock Broad lifetime distributions for ordering dynamics in complex
  networks.
\newblock {\em Physical Review E}, 2008.
\newblock (in press) [e-print arXiv:0808.3318].

\bibitem{Baronchelli_ng_long}
A.~Baronchelli, V.~Loreto, and L.~Steels.
\newblock In-depth analysis of the naming game dynamics: the homogeneous mixing
  case.
\newblock {\em Interntaional Journal of Modern Physics C}, 19:785, 2008.

\bibitem{Baronchelli_2005}
A.~Baronchelli, L.~Dall'Asta, A.~Barrat, and V.~Loreto.
\newblock Topology induced coarsening in language games.
\newblock {\em Physical Review E}, 73:015102, 2005.

\bibitem{dallasta06}
L.~Dall'Asta, A.~Baronchelli, A.~Barrat, and V.~Loreto.
\newblock Agreement dynamics on small-world networks.
\newblock {\em Europhysics Letters}, 73(6):969--975, 2006.

\bibitem{dallasta06b}
L.~Dall'Asta, A.~Baronchelli, A.~Barrat, and V.~Loreto.
\newblock Nonequilibrium dynamics of language games on complex networks.
\newblock {\em Physical Review E}, 74(3):036105, 2006.

\bibitem{dallasta06c}
L.~Dall'Asta and A.~Baronchelli.
\newblock Microscopic activity patterns in the naming game.
\newblock {\em Journal of Physics A: Mathematical and General},
  39:14851--14867, 2006.

\bibitem{Baronchelli_2007}
A.~Baronchelli, L.~Dall'Asta, A.~Barrat, and V.~Loreto.
\newblock Nonequilibrium phase transition in negotiation dynamics.
\newblock {\em Physical Review E}, 76(5):051102, 2007.

\bibitem{Gunton_1983}
J.~D. Gunton, M.~San~Miguel, and P.~Sahni.
\newblock {\em Phase Transitions and Critical Phenomena}, volume~8, chapter The
  dynamics of first order phase transitions, pages 269--446.
\newblock Academic Press, London, 1983.

\end{thebibliography}
\end{document}